\documentclass[referee]{raa}            
\usepackage{graphicx,times}             
\input{epsf.sty}                        
\input{psfig.sty}                       

\begin{document}

\title{96-antenna radioheliograph}

\volnopage{Vol.0 (200x) No.0, 000--000}      
\setcounter{page}{1}          

\author{S.V. Lesovoi\inst{1} \and A.T. Altyntsev\inst{1} \and E.F. Ivanov\inst{1} \and A.V. Gubin\inst{1}}

\institute{Institute of Solar-Terrestrial Physics, Irkutsk, Russia; {\it svlesovoi@gmail.com}}

\date{Received~~2014 month day; accepted~~2014~~month day}

\abstract{Here we briefly present some design approaches for a multifrequency 96-antenna radioheliograph. The array antenna configuration, transmission lines and digital receivers are the main focus of this work. The radioheliograph is a T-shaped centrally-condensed radiointerferometer operating at the frequency range 4-8~GHz. The justification for the choice of such a configuration is discussed. The antenna signals are transmitted to a workroom by analog optical links. The dynamic range and phase errors of the microwave-over-optical signal are considered. The signals after downconverting are processed by the digital receivers for delay tracking and fringe stopping. The required delay tracking step and data rates are considered. Two 3-bit data streams (I and Q) are transmitted to a correlator with the transceivers embedded in FPGA (Field Programmed Gate Array) chips and with PCI Express cables.}

\authorrunning{S.V. Lesovoi, A.T. Altyntsev, E.F. Ivanov \& A.V. Gubin}            
\titlerunning{96-antenna radioheliograph }  

\maketitle
\section{Introduction}
\label{sect:intro}
The development of the next-generation, ground-based solar radio telescopes is aimed at the solution of current problems of solar-terrestrial physics. A few next generation radioheliographs are being under construction. The main objective of such instruments is imaging the Sun with spatial and spectral resolution simultaneously. Several instruments are now under construction. Chinese Spectral Radioheliograph~(CSRH) is being constructed in Inner Mongolia of China (Yan et al.~\cite{Yan2011}, Yan et al.,~\cite{Yan2013}). CSRH-I (0.4-2.0~GHz) is now in test mode, and CSRH-II (2-15~GHz) is mounted. The Expanded Owen Valley Solar Array was operated in the frequency range 1-9~GHz in order to test design elements (Gary et al.~\cite{Gary2012}) and is now nearing completion with operation in the range 2.5-18~GHz. The 96-antenna multifrequency radioheliograph is under construction now. This radioheliograph is upgrade of the Siberian Solar Radio Telescope (SSRT) (Grechnev et al.~\cite{Grechnev2003}). The radioheliograph design approaches were discussed in the description of the 10-antenna prototype (Lesovoi et al.~\cite{Lesovoi2012}). Here we focus on some of the key features that need to be emphasized.  In this work we would like to consider the following issues: the antenna configuration, the dynamic range and phase errors of the analog optical links and the delay tracking step requirement.

\section{Antenna configuration}
\label{sect:array}
Because the existing antenna stations of the SSRT are used for the radioheliograph antennas, the antenna configuration of the radioheliograph is T-shaped centrally-condensed array. The shortest baseline is 4.9~m, while the longest are 622.30~m and 311.15~m in the East-West and South directions. The variable distance between adjacent antennas is determined not only by consideration of the resulting point spread function (PSF) but also the requirement to fit 32 antennas in the each arm of the T-shaped array. The total number of baselines of the 96-antenna radioheliograph is 4560. The Figure~1 shows the distribution of 2512 1-dimensional baselines along the East-West and South directions. These baselines are redundant except for four baselines of South array and for eleven baselines of East-West array. The responses of these baselines on the solar microwave signal will be used to calibrate the antenna phases and magnitudes. The total number of non-redundant 2-dimensional baselines to imaging the Sun is of 2048.
\label{sect:receivers}
\begin{figure}
	\label{Fig:antConf}
	\centering
	\includegraphics[width=8cm, angle=0]{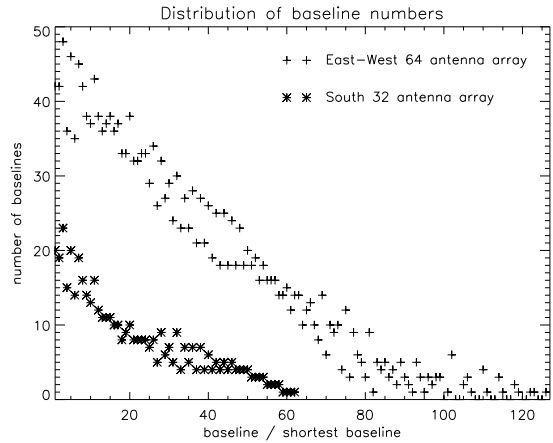}
	\caption{Distribution of 1-dimensional baselines for the South and East-West antenna arrays of the radioheliograph. With the exception of four baselines for the South array and eleven for the East-West array all baselines are redundant.}
\end{figure}
The East-West cross-section of the 2-dimensional radioheliograph PSF is shown on the Figures~2,~3. It is distinctly seen from the Figures~2,~3 that the sidelobe levels of the non-uniform (centrally-condensed) and uniform arrays are comparable only in the vicinity of the main lobe. The high level of the far sidelobes is one of reasons to use the uniform array, because more precision is required to calculate far sidelobes in the dirty PSF. Although the uniform array has better spatial resolution and, especially, distant sidelobe performance than the centrally-condensed array, nevertheless the added expense is prohibitive.

\begin{figure}[h]
	\begin{minipage}[t]{0.495\linewidth}
		\centering
		\includegraphics[width=5cm, angle=0]{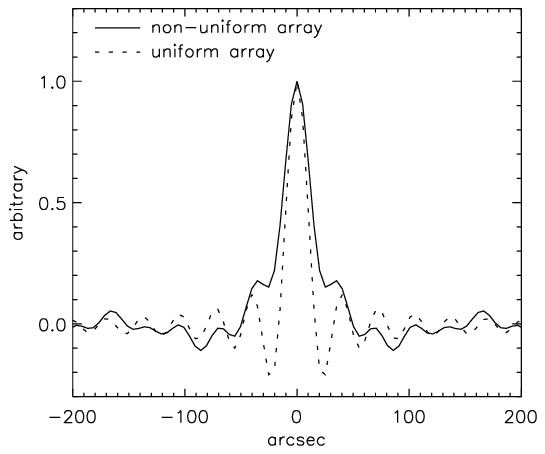}
		\caption{Point spread functions of both non-uniform and uniform arrays. Sidelobe levels are comparable in the vicinity of the main lobe.}
		\label{Fig:PSF1}
	\end{minipage}
	\begin{minipage}[t]{0.495\linewidth}
		\centering
		\includegraphics[width=5cm, angle=0]{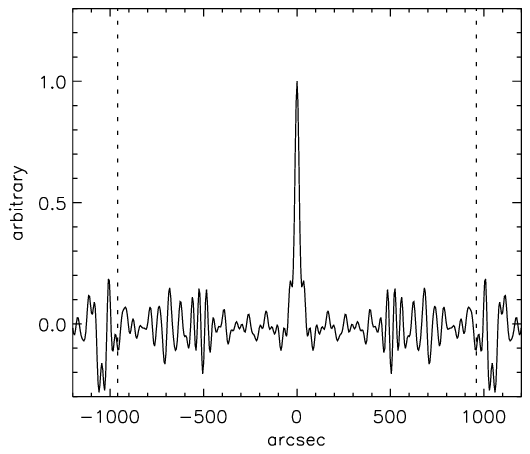}
		\caption{Point spread function of the non-uniform array. Dashed lines show the angular size of the Sun. Sidelobe level significantly increases with distance from the main lobe. }
		\label{Fig:PSF2}
	\end{minipage}
\end{figure}

\section{Analog fiber optical links}
\label{sect:links}
An analog optical link with direct modulation of a laser diode is used for transmission of the RF signal from an antenna to a workroom. The compensation of the high losses of the analog optical link is provided by using a RF preamplifier. But the dynamic range and the phase errors in the RF signal should be considered. The typical spur free dynamic range~(3rd order) of the analog optical link is about SFDR$_3$=100~dB~Hz$^{2/3}$. The dynamic range for a bandwidth of 4~GHz will be SFDR$_{3\Delta f}$ = SFDR$_3$ - $2/3\cdot10\cdot\lg(4\cdot10^9)\approx$36~dB. The fluxes of the quiet Sun are about 200~s.f.u. for the operating frequency range 4-8~GHz.  So, it is possible to observe the microwave bursts with the peak flux up to 4$\cdot$10$^5$ s.f.u. Because the occurrence of such high peak flux bursts is negligible (Nita et al.~\cite{Nita2002}), it is not necessary to use a switched attenuator at the antenna front end for solar flares observation.

The sources of the phase errors in the RF signal after its conversion from optical to electrical form are the finite spectral width of the laser emission and chromatic dispersion of the fiber. The semiconductor single mode laser spectral width is about 0.5~nm. The typical chromatic dispersion of the fiber near wavelength 1310~nm is about 1~ps/(km~$\cdot$~nm). The length of the optical link in our design is about 0.4~km, which leads to a delay error of 0.2~ps. The phase error at the frequency of 6~GHz will be about 0.5$^\circ$, that is acceptable. The temperature dependence of the transmission delay also leads to the phase error at microwaves. The typical temperature drift of the delay is about 40-130~ps/(km~$\cdot$~C$^\circ$). The local diurnal temperature change is about 20~C$^\circ$ and the fact that only about 5~m optical cable lays outdoor result as diurnal delay change about 10~ps. The phase change at the frequency 6~GHz is about 20$^\circ$ per day. The correlation is affected not by the phase at each antenna but by the differences of these phases. One can expect these differences will be much less relative of phase errors, i.e. about 2$^\circ$ per day, that is acceptable.

\label{sect:receivers}
\begin{figure}
	\label{Fig:digRec}
	\centering
	\includegraphics[width=10cm, angle=0]{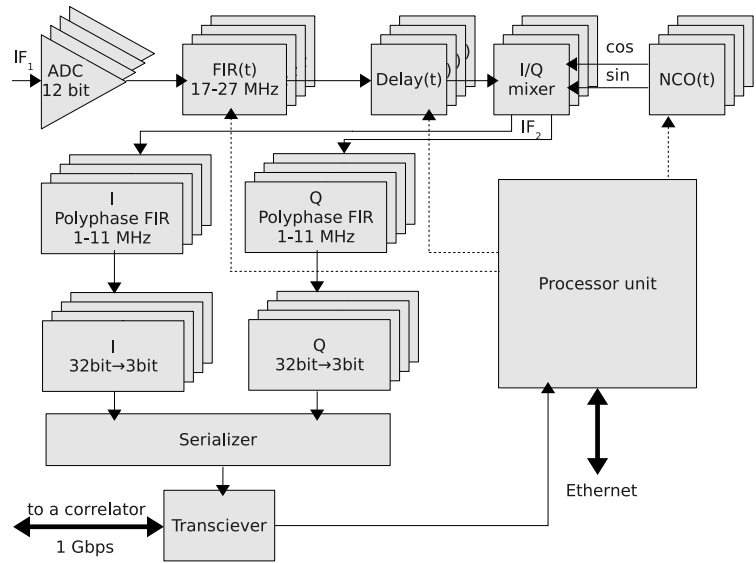}
	\caption{Digital receiver module for four antennas. Two 3-bit streams (I, Q) are transmitted to a correlator with the transceiver in Basic functional mode.}
\end{figure}

\section{Digital receivers}
Each digital receiver module (Figure~4) of the radioheliograph processes signals from four antennas. It converts first intermediate frequency (IF$_1$) signals to digital data and forms the passband 17-27~MHz. The 64-th order finite impulse response filter (FIR) is used both to form the passband and for fractional-sample delay. The digital downconvertor is used simultaneously to produce IQ signals, to fringe stopping and to downconvert IF$_1$ signals to a second intermediate frequency passband 1-11~MHz (IF$_2$). The data rate required for transmission from the digital receiver to a correlator can be two times less due to such IF$_2$ passband. So, the instantaneous frequency bandwidth (i.e. spectral resolution) of the radioheliograph is 10~MHz. The number of frequency channels is limited only by the desired temporal resolution. Because the integration time is about 0.1~s, the temporal resolution for the case of 100 frequency channel is about 10~s. The delay errors result in the IF$_2$ signal phase slope error $2\pi\nu_{if2}(\delta\tau_g^k - \delta\tau_g^l)=2\pi\nu_{if2}\Delta\tau_g^{kl}$, where $\delta\tau_g^k = \tau_g^k-n\tau_{step}$ and $\delta\tau_g^l=\tau_g^l-m\tau_{step}$ are delays errors for $k ,l$ antennas, and $\tau_{step}$ is the discrete step of the instrumental delay. The rms frequency within the passband is $\Delta\nu_{if2} / 2 \sqrt{3}$. Because the rms delay error is $\tau_{step}/6$ (Thompson~\cite{Thompson2001}), one can infer that phase error is $2\pi \frac{\Delta\nu_{if2}} {2 \sqrt{3}} \frac{\tau_{step}}{6} \leq 1^\circ$ for the delay step of 1~ns. But the delay errors lead to significant displacements of a two-element interferometer beam pattern with respect to the single dish one: $\Delta\theta = c\Delta\tau_g^{kl}/(|\vec{b^k} - \vec{b^l}|\sin{\theta})$, where $\vec{b^k},~\vec{b^l}$ are baselines of $k, l$ antennas, $c$ is speed of light, $\theta$ is the source angle. The angular displacement for a delay step of 1~ns for the shortest baseline at culmination is $\Delta\theta \approx 3.5^\circ$. This is excessive, therefore the step of the fractional-sample delay is chosen as 0.1~ns. In practice, such a small delay step is sufficient to smooth phase jumps in correlations on the shortest baselines.
The 3-bit I and 3-bit Q digital receiver data to be transmitted to the correlator. The sample rate of the data is 25~MHz. The whole receiver net data rate is 600~Msps and data rate with payload bits is 800~Msps. Finally, the correlator input data rate after 8b/10b encoding shall be 24~Gbps for 96 antennas. This data stream can be transmitted from the receiver modules to the correlator via transceivers embedded in Altera\texttrademark FPGA (Field Programmed Gate Array) chips. The PCI Express over cable solution is used. More specifically, PCI Express cables and the transceivers in Basic functional mode are used. This solution allows calculation of all complex correlations of the 96-antenna interferometer with a single FPGA chip.

\section{Conclusions}
The antenna configuration, analog optical links and the digital receivers of the 96-antenna radioheliograph are considered. The T-shape antenna configuration is chosen just because the existing antenna stations of the SSRT are used for the radioheliograph. The centrally-condensed array is chosen because it is less expensive than an equivalent uniform array. It is shown that the spur free dynamic range of the analog optical links is sufficient to observe the solar flares without using a switched attenuator. The phase errors and phase temperature drifts of the link are acceptable. Because of the short baselines (a few meters), typical for solar radiointerferometers, the delay tracking step of 0.1~ns is needed. The practice shows that such step can be realized with the fractional-sample delaying.
\label{sect:conclusion}

\begin{acknowledgements}
We thank anonymous referee for constructive comments.
This study was supported by the
Russian Foundation of Basic Research (12-02-91161, 12-02-00173, 12-02-10006, 13-02-90472) and by a Marie Curie International
Research Staff Exchange Scheme Fellowship within the
7th European Community Framework Programme. The
work is supported in part by grants of Ministry of education
and science of the Russian Federation (State Contracts
16.518.11.7065 and 02.740.11.0576).
\end{acknowledgements}

\label{lastpage}

\end{document}